\documentclass[preprints,article,submit,pdftex,moreauthors]{Definitions/mdpi} 

\usepackage{graphicx}
\usepackage{overpic}
\usepackage{subcaption}
\preto{\abstractkeywords}{\nolinenumbers}

\firstpage{1} 
\pubvolume{1}
\issuenum{1}
\articlenumber{0}
\pubyear{2026}
\copyrightyear{2026}
\datereceived{ } 
\daterevised{ } 
\dateaccepted{ } 
\datepublished{ } 

\Title{Influence of compaction pressure on the impedance of Gadolinium Doped Ceria electrolytes for IT-SOFCs}

\Author{Renato A. N. de Oliveira \textsuperscript{1}*, 
            Celeste Z.A. Aquino\textsuperscript{4},
            Magna Monteiro Schaerer\textsuperscript{4},
            Rafael Galiza Yoshimura\textsuperscript{1},
            Liliana Mogni\textsuperscript{2}, 
            Maurico Arce\textsuperscript{2},
            Diego G. Lamas\textsuperscript{3},
            Lucia Toscani\textsuperscript{3},
            Maria Belén Arcentales Vera\textsuperscript{3},
            Esteban Elvis Asto Ramos\textsuperscript{3},
            Marcia Carvalho de Abreu Fantini\textsuperscript{5},
            Taofeeq Oladayo Bello\textsuperscript{5},
            Tereza da Silva Martins\textsuperscript{6},
            Danilo Waismann Losito\textsuperscript{6},
            James Moraes de Almeida\textsuperscript{7},
            Valeria Spolon Marangoni\textsuperscript{7},
            Leopoldo Suescunand\textsuperscript{8}, 
            and Santiago Figueroa \textsuperscript{9}* }

\AuthorNames{}

\address{%
\textsuperscript{1} \quad LNLS/CNPEM-Brasil \\
\textsuperscript{2} \quad CAB-Argentina \\
\textsuperscript{3} \quad UNSAM-Argentina \\
\textsuperscript{4} \quad UNA-Paraguay \\
\textsuperscript{5} \quad USP-Brasil \\
\textsuperscript{6} \quad UNESP-Brasil \\
\textsuperscript{7} \quad Ilum-Brasil \\
\textsuperscript{8} \quad UdelaR-Uruguai \\
\textsuperscript{9} \quad Unicamp-Brasil \\
}

\corres{Correspondence: renato.oliveira@lnls.br;santiago.figueroa74@gmail.com}

\abstract{ 
Gadolinium doped ceria (GDC) is one promising oxygen‑ion conducting ceramic electrolyte for intermediate‑temperature solid oxide fuel cells (IT‑SOFCs), due to its high ionic conductivity at reduced operating temperatures, favorable defect chemistry, and compatibility with a broad range of electrode materials \cite{skinnerd:2008, Saddam:2020}. Despite extensive understanding of its intrinsic ion transport mechanisms, the influence of ceramic processing parameters on the effective electrical behavior of polycrystalline GDC electrolytes remains an active topic for investigation \cite{GDC_LSGM_DATA,Acharya17022022,JiaHong2025}. In particular, processing steps that govern green body formation and sintering can strongly affect microstructural features such as density, grain size, grain boundary character, and residual porosity, which in turn determine the macroscopic conductivity \cite{Acharya17022022,JiaHong2025,Liang2022}. In this work, dense GDC ceramic pellets were fabricated under systematically varied isostatic compaction pressures ranging from 49 to 140 MPa, followed by sintering at 1350 \textsuperscript{$\circ$}C for four hours under identical thermal conditions. Platinum electrodes were deposited on both sides of the pellets by electron‑beam deposition, and the electrical properties were characterized by electrochemical impedance spectroscopy (EIS) over a wide temperature range. The results demonstrate a dependence of the impedance with respect to compaction pressure.}

\begin{document}
\nolinenumbers

\keyword{Solid oxide fuel cells} 


\section{Introduction}

Solid oxide fuel cells are electrochemical energy conversion devices capable of converting the chemical energy of fuels directly into electrical energy with high efficiency \cite{skinnerd:2008,Saddam:2020,Fangjie:2025}. Central to SOFC operation is the solid electrolyte, which must support fast oxygen‑ion transport while remaining electronically insulating and chemically stable across a wide range of temperatures and oxygen partial pressures \cite{Unicamp2025,Liliana2016}. Among the oxide‑ion conducting ceramics investigated to date, gadolinium doped ceria (Ce$_{1-x}$Gd$_{x}$O$_{2-\delta}$, GDC) has emerged as a candidate for intermediate‑temperature operation, typically between 500 and 700 \textsuperscript{$\circ$}C \cite{Liang2022,Vinchhi2024}. 

The crystal structure of ceria provides a highly symmetric oxygen sublattice that is particularly conducive to vacancy‑mediated ionic transport \cite{batista2016,GHAEMI2024}. Aliovalent substitution of Ce\textsuperscript{4+} by Gd\textsuperscript{3+} introduces oxygen vacancies to maintain charge neutrality, thereby increasing the concentration of mobile defects responsible for oxide‑ion conduction \cite{MOMIN2022,MinalGupta2022}. When the dopant ionic radius closely matches that of the host cation, as is the case for Gd\textsuperscript{3+} in CeO\textsubscript{2}, the vacancy mobility is enhanced \cite{COSTILLAAGUILAR2021}. As a result, GDC exhibits significantly higher ionic conductivity than conventional yttria‑stabilized zirconia (YSZ) at temperatures below 800 \textsuperscript{$\circ$}C, enabling reduced SOFC operating temperatures and mitigating degradation phenomena such as electrode inter-diffusion, thermal expansion mismatch, and accelerated material aging.

Beyond its favorable transport properties, GDC offers additional advantages for SOFC applications, including relatively low sintering temperatures, chemical compatibility with a variety of electrode materials, and tolerance to redox cycling \cite{WANG2022,CABRERAPASCA2026}. These attributes have motivated extensive research into GDC‑based electrolytes, interlayers, and composite electrodes for IT‑SOFC systems \cite{CABRERAPASCA2026}.

While the intrinsic defect chemistry and bulk transport mechanisms of GDC are well established, the effective ionic conductivity measured in polycrystalline ceramics is strongly influenced by microstructural features introduced during powder processing and sintering. In dense ceramics, the total electrical response arises from contributions of both grain bulk, and grain boundaries \cite{THIAGO2026}. Grain boundaries increase resistivity due to space‑charge effects, and structural inhomogeneities. Consequently, the nature of grain boundaries play a critical role in determining macroscopic conductivity \cite{Zhou2002}.

Ceramic processing parameters such as powder characteristics, green body density, compaction pressure, and sintering conditions directly affect porosity and grain boundary area. Compaction pressure during pellet formation governs the initial density and homogeneity of the powder pellet, influencing sintering kinetics and the microstructure evolution.

The objective of the present study is to investigate how the impedance of GDC can be influenced through controlled variation of the compaction pressure applied during sample preparation. Holding all sintering parameters constant and varying only the isostatic pressing pressure, the role of compaction pressure on the resulting electrochemical behavior is isolated. Electrochemical impedance spectroscopy (EIS) is employed as the primary characterization tool to probe the electrical properties of GDC as a function of temperature and to elucidate the underlying conductivity mechanisms. This approach enables separation of bulk and grain‑boundary contributions and provides insight into how processing induced microstructural changes affect charge transport.

\section{Materials and Methods}

\begin{figure}[h]
\centering  
    \begin{overpic}[width=1.0\linewidth,keepaspectratio,angle=0,trim=0.cm 0.cm 0.cm 0.cm, clip]{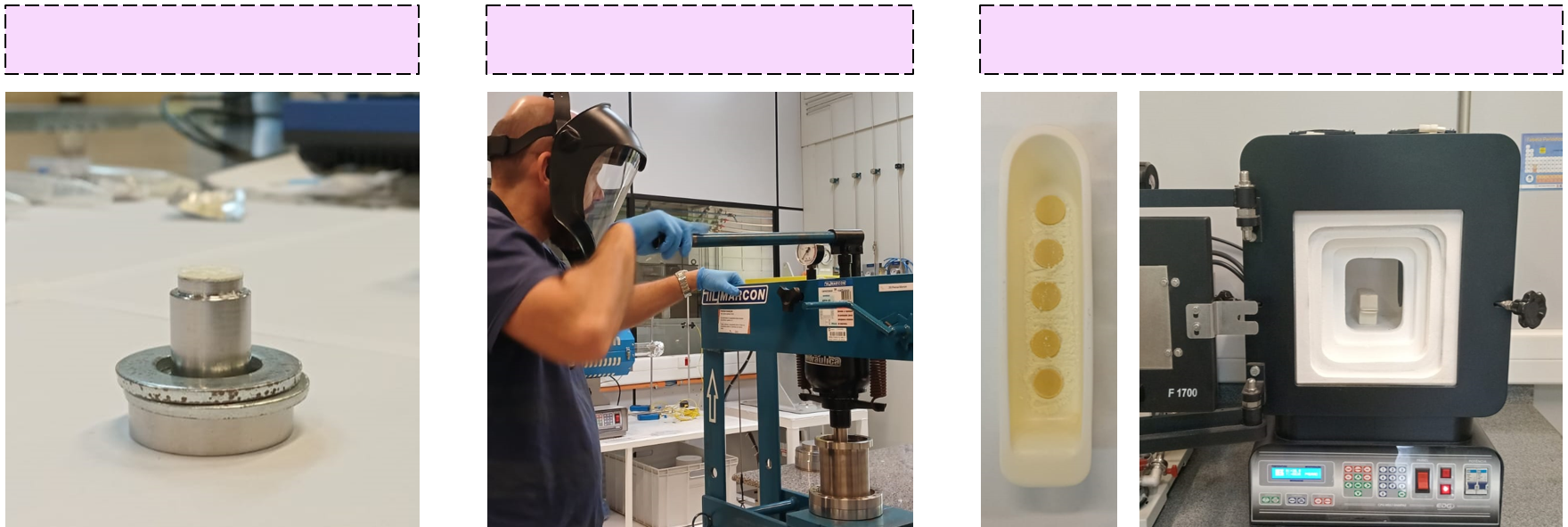}
    \put(1,30.5){\textcolor{blue}{1. Pellet conformation}}
    \put(31.75,30.5){\textcolor{blue}{2. Isostatic pressing}}
    \put(63.25,30.5){\textcolor{blue}{3. Sintering}} 
    \end{overpic}
    \caption{Fabrication steps for GDC ceramic pellets.}
    \label{LNNANOsetup}
\end{figure}

Commercially available gadolinium doped ceria powder with nominal composition Ce$_{0.8}$Gd$_{0.2}$O$_{2-\delta}$ (from  Fuel Cell Materials Inc., OH, USA) was processed at the Laboratory for Nano Ceramic Processing at LNNano/CNPEM \cite{LNNANOSITE}. The powder was weighed prior to compaction to ensure reproducible packing behavior. As shown in Fig. \ref{LNNANOsetup}, green pellets were fabricated using an in‑house‑developed isostatic pressing device, which applies uniform hydrostatic pressure to minimize density gradients. Compaction pressures were varied systematically from 49 to 140 MPa, producing a series of pellets with identical geometry (see Fig. \ref{LNNANOsetup}).

After compaction, all pellets were sintered in air at 1350 \textsuperscript{$\circ$}C for four hours. To minimize thermal stresses, heating and cooling rates were controlled and set to 5 \textsuperscript{$\circ$}C per minute. This sintering procedure was selected to achieve high densification of GDC while allowing grain‑boundary effects to remain discernible in subsequent electrical measurements.

Following sintering, the density of each pellet was measured using the Archimedes hydrostatic method. Measurements were performed with a hydrostatic density meter, enabling accurate determination of bulk density. These density measurements provided a quantitative basis for correlating compaction pressure, densification, and electrical conductivity.

\begin{figure}[h]
\centering
    \includegraphics[width=0.35\linewidth,trim=1.cm 5.cm 2.cm 8.5cm, clip]{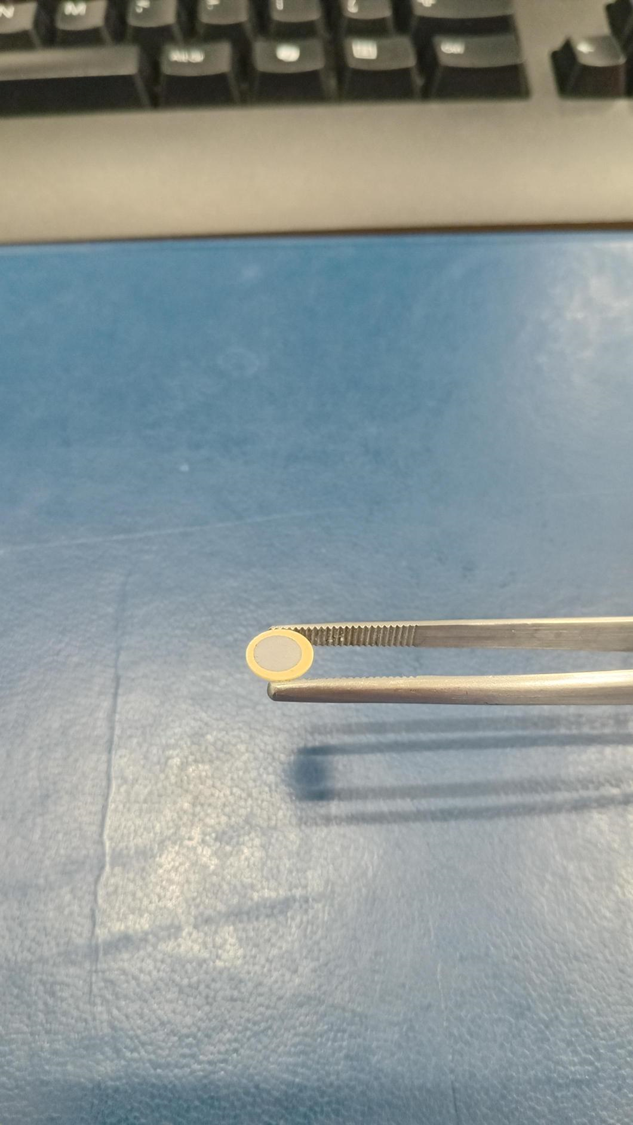}
    \caption{9 $mm$ diameter and 600 $\mu m$ thick GDC ceramic pellet. Circular platinum electrodes with a diameter of 6 $mm$ and approximately 0.2 $\mu m$ thick were deposited in the center on both faces of the pellet by electron‑beam deposition.}
    \label{pellet}
\end{figure}

After density characterization, circular platinum electrodes with a diameter of 6 $mm$ and approximately 0.2 $\mu m$ thick were deposited in the center on both faces of each pellet by electron‑beam deposition (see Fig. \ref{pellet}). Platinum was selected due to its high electronic conductivity, chemical inertness, and thermal stability over the investigated temperature range. The thin, dense electrodes ensured reproducible electrical contact and minimized electrode polarization effects.

\begin{figure}
\centering
    \includegraphics[width=.55\linewidth]{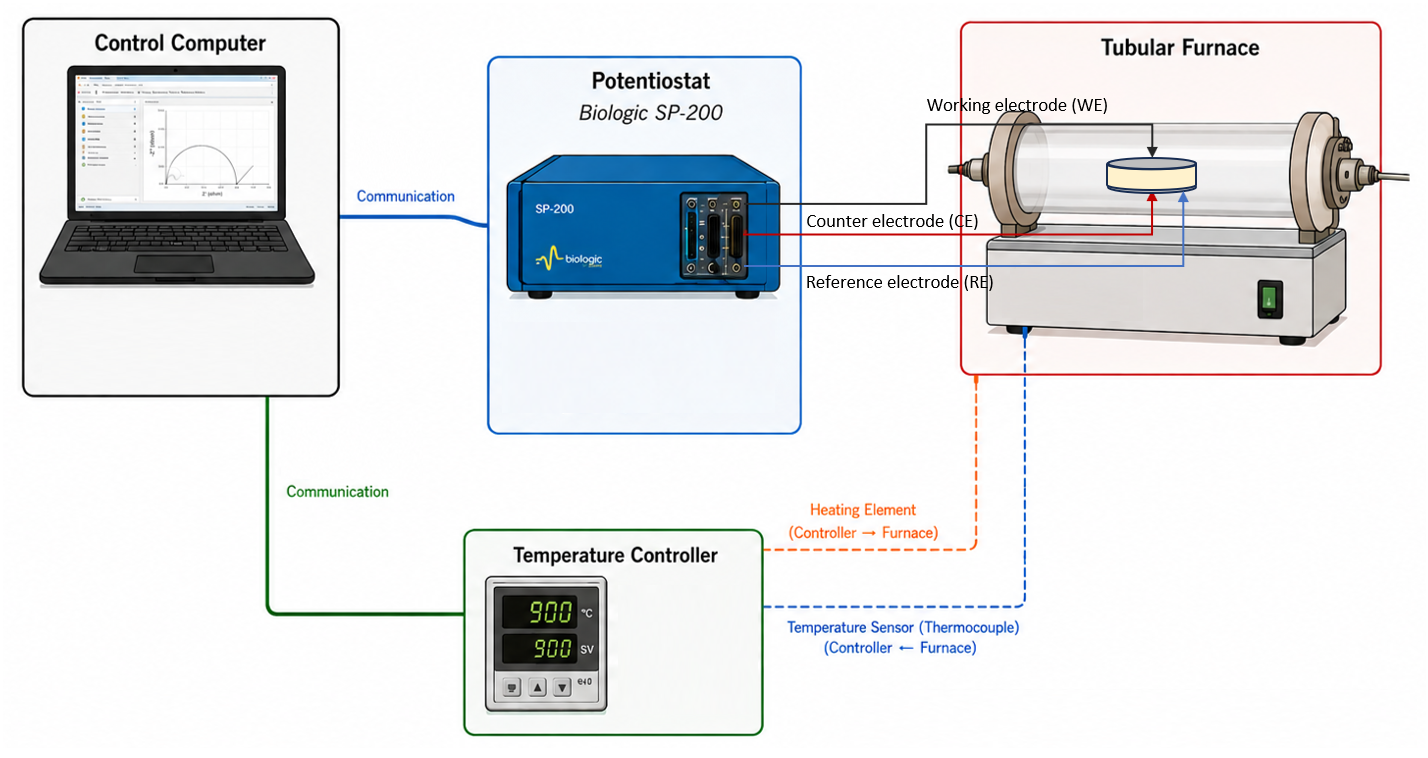}
    
    \begin{overpic}[width=.40\linewidth,keepaspectratio,angle=270,trim=10.cm 0.cm 50.cm 10.cm, clip]{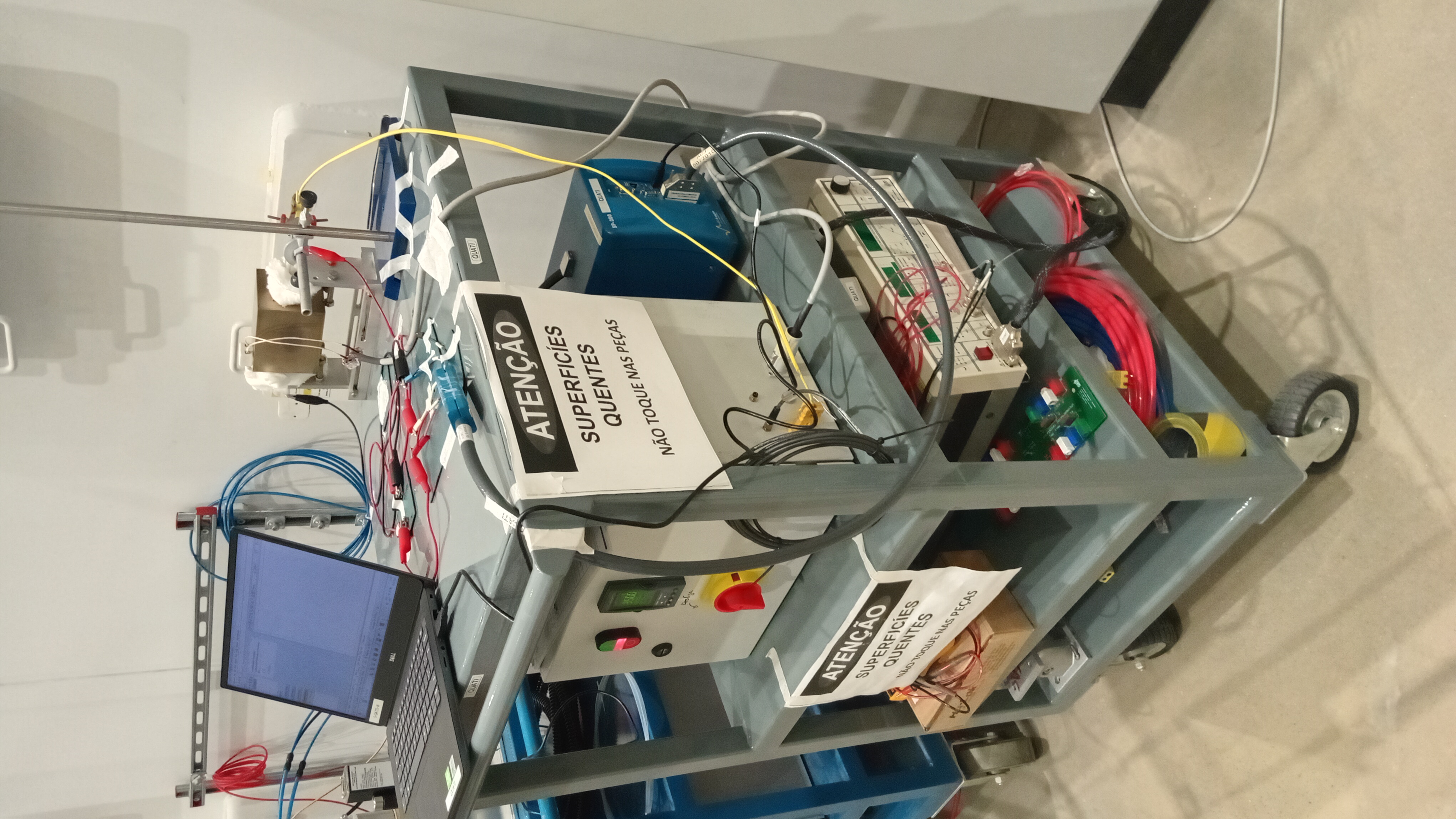}
    \put(-15,15){\textcolor{blue}{Temperature Controller}}
    \put(80,35){\textcolor{blue}{Potentiostat}}
    \put(45,87){\textcolor{blue}{Tubular Furnace}}
    \put(-20,70){\textcolor{blue}{Control Computer}}
    \end{overpic}
    
    \caption{(Top) Schematic diagram of the setup commissioned at Sirius/Quati beam line for the impedance spectroscopy measurements. (Bottom) Picture showing the setup on the experimental hutch.}
    \label{setupPEIS}
\end{figure}

Electrochemical impedance spectroscopy (EIS) measurements were performed using an SP-200 Biologic potentiostat commissioned at the Sirius/QUATI beamline \cite{Santiago2023} and dedicated to in-situ impedance spectroscopy experiments (see Fig. \ref{setupPEIS}). The platinum electrodes on opposite sides of the pellet served as the working and counter electrodes, with electrical contact established via platinum wires. The sample assembly was placed inside an in‑house‑developed furnace capable of precise temperature control and stabilization within ±0.1 \textsuperscript{$\circ$}C across the sample volume. The platinum electrodes on opposite sides of the pellet served as the working and counter electrodes, with external electrical contact established via nickel foam collectors and platinum wires on both sides.

The electrochemical impedance spectra were collected, at multiple temperatures, using the potentiostatic electrochemical impedance spectroscopy technique, applying a sine wave as the excitation signal with frequencies ranging from 1MHz to 1Hz with 15 points per decade, amplitude of 100mV, and under open circuit voltage conditions. The impedance measurements were performed twice: the first measurement was acquired after a stabilization period of 1 h, and the second after an additional 0.5 h. The measurements were considered valid only when no significant discrepancies were observed between the two datasets.

\section{Results}

\begin{figure}
    \centering
    
    \begin{overpic}[width=.35\linewidth]{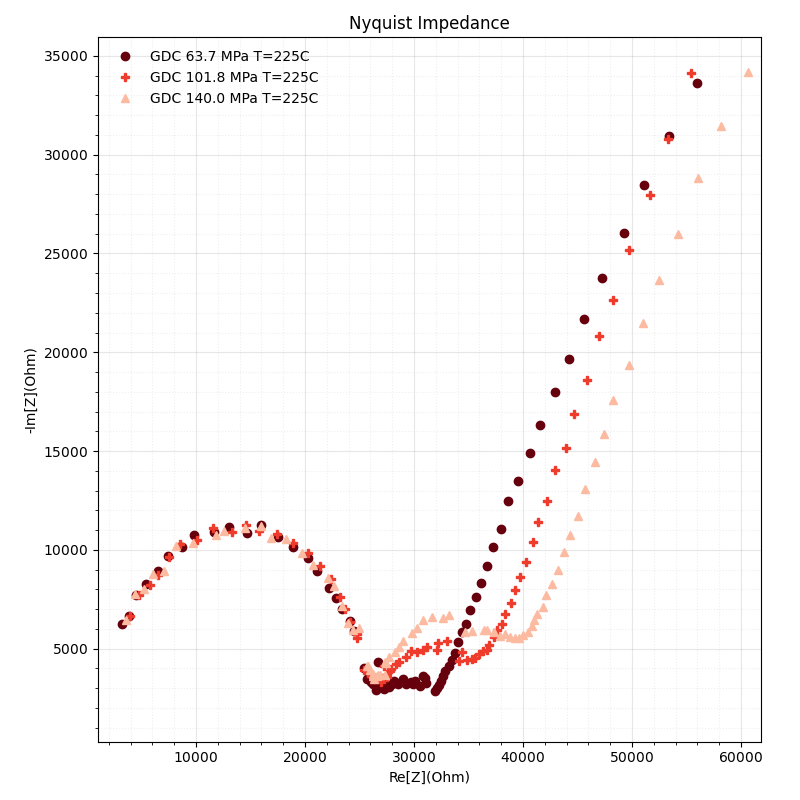}
    \put(22,46){\textcolor{black}{\scriptsize Bulk}}
    \put(25,38){\Large \bf \textcolor{black}{ $\downarrow$ }}
    
    \put(70,19){\textcolor{black}{\scriptsize Grain}}
    \put(70,14){\textcolor{black}{\scriptsize boundary}}
    \put(58,16){\Large \bf \textcolor{black}{ $\leftarrow$ }}
    
    \put(45,55){\textcolor{black}{\scriptsize Electrode}}
    \put(65,54){\Large \bf \textcolor{black}{ $\rightarrow$ }}
    \end{overpic}    
    \includegraphics[width=.35\linewidth]{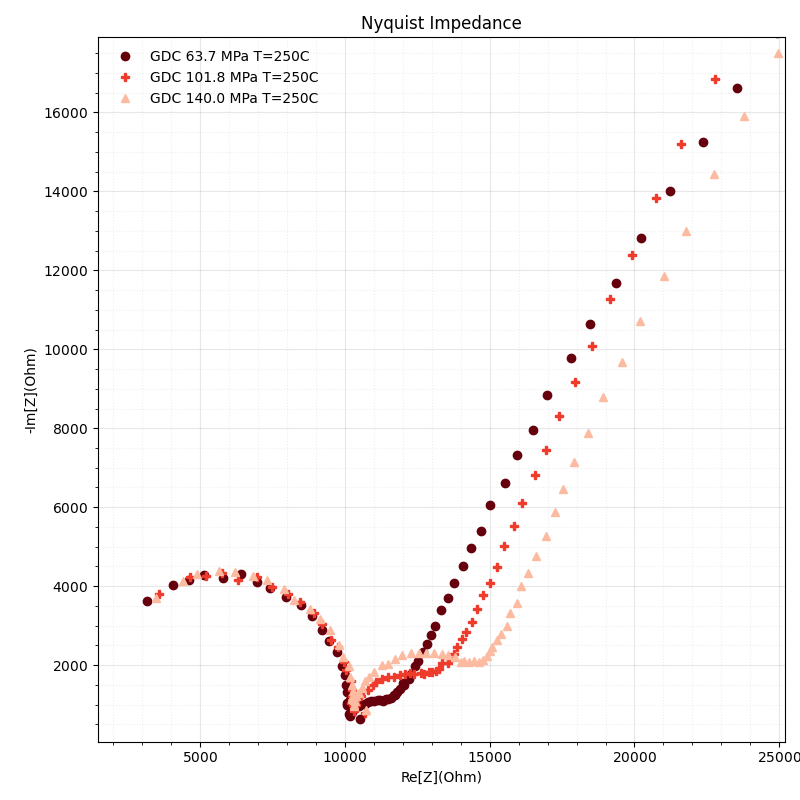}
    \includegraphics[width=.35\linewidth]{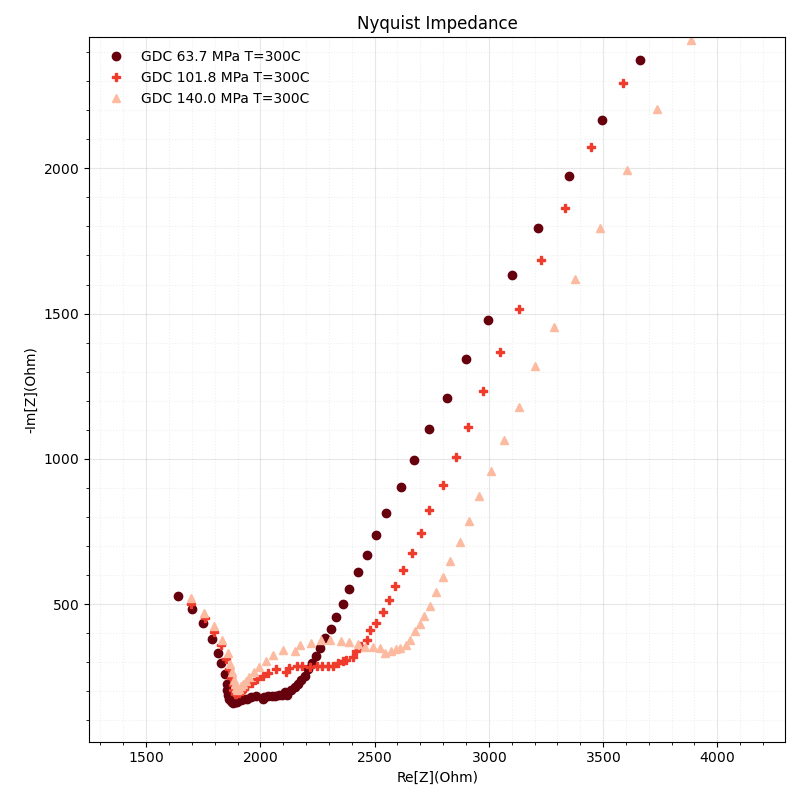}
    \caption{Nyquist plots for T=225 \textsuperscript{$\circ$}C, 250 \textsuperscript{$\circ$}C and 300 \textsuperscript{$\circ$}C. All plots shown two partially overlapping semicircular arcs followed by a low-frequency tail related to electrode contribution.}
    \label{impedance_results}
\end{figure}
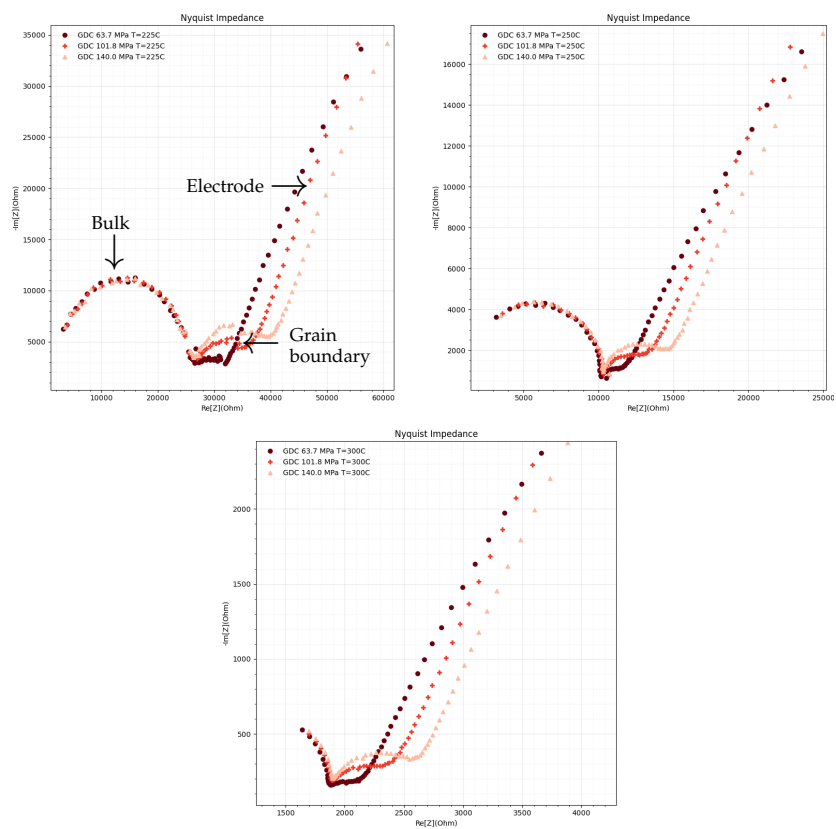

Electrochemical impedance spectroscopy measurements were carried out in air over a temperature range from 200 \textsuperscript{$\circ$}C to 350 \textsuperscript{$\circ$}C, for all GDC pellets produced under different isostatic compaction pressures. Nyquist plots exhibit the characteristic response of dense oxygen-ion conducting ceramics, consisting of two partially overlapping semicircular arcs followed by a low-frequency electrode-related contribution (see Fig. \ref{impedance_results}). The high-frequency semi-circle corresponds to the bulk resistance, whereas the intermediate-frequency arc is attributed to grain-boundary contributions. The low-frequency tail, which becomes more pronounced at elevated temperatures, is associated with electrode polarization.

Changes in the impedance spectra with compaction pressure are observed. Pellets compacted at lower pressures, show relatively lower grain-boundary arcs, indicating lower resistive contributions from intergranular regions. As the compaction pressure increases toward intermediate and high values, an increase in the grain-boundary resistance is observed manifested primarily as an enlargement of the grain-boundary arc. This behavior demonstrates that compaction pressure influences the sample grain-boundary impedance leaving the bulk resistance unaffected.

\begin{figure}[h]
\centering
    \includegraphics[width=0.35\linewidth,trim=0.cm 0.cm 0.cm 0.cm, clip]{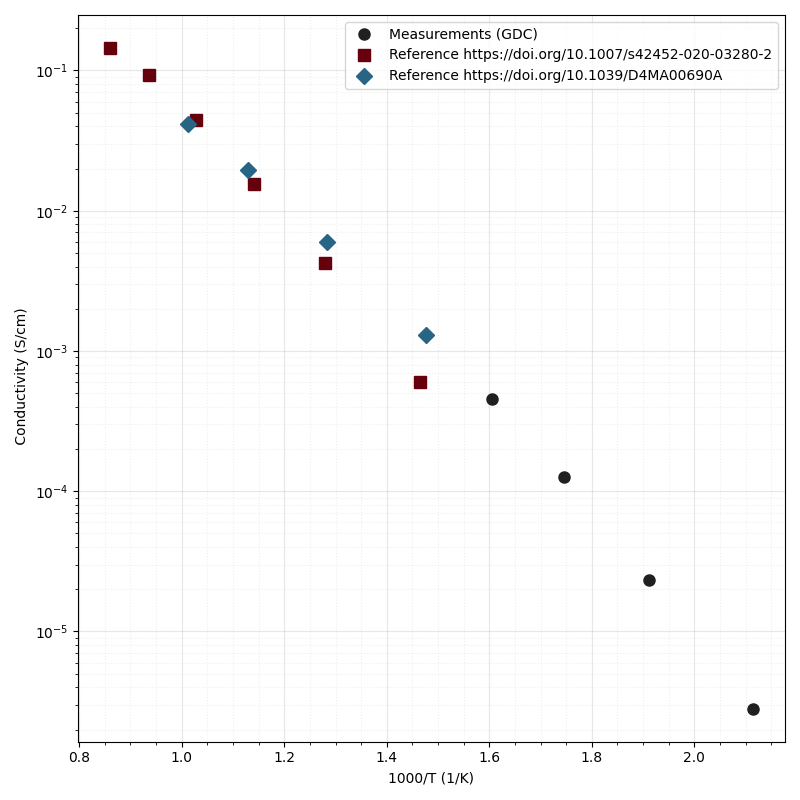}
    \caption{Arrhenius plot for gadolinium doped ceria (GDC). The results were compared with previously reported measurements extracted from the literature \cite{GDC_DATA1,GDC_LSGM_DATA}.}
    \label{Arrhenius}
\end{figure}

With the bulk resistance values extracted from the measurements, and the sample pellet geometry, the bulk conductivity was calculated as a function of temperature, as shown in the Arrhenius plot in Fig. \ref{Arrhenius}. Furthermore, the results obtained in this work were compared with previously reported measurements from the literature \cite{GDC_DATA1,GDC_LSGM_DATA}, demonstrating good agreement between the datasets.

\begin{figure}
    \centering
    \includegraphics[width=.35\linewidth]{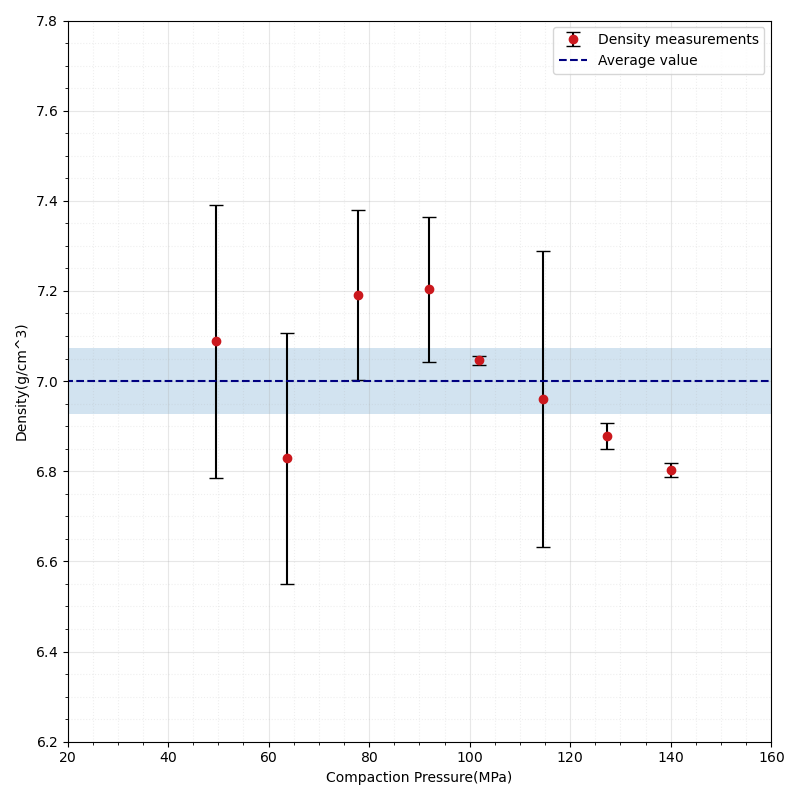}
    \caption{Measured density versus compaction pressure. The dashed line indicates the average value calculated from the measured data points, and the blue area the calculated error for the average value considering the experimental error bars on the data points.}
    \label{density}
\end{figure}

Furthermore, post sintering density measurements performed using the Archimedes method reveal a constant relation between compaction pressure and final density as shown in Fig. \ref{density}. Considering the experimental error, an overall plateau with respect to compaction pressure is observed from 40 to 140 MPa, resulting in an average density value of  7.00 $\pm$ 0.07 g/cm$^3$. 

Finally, based on the available measurements, the grain-boundary contribution for the observed variations in impedance with compaction pressure is hypothetically attributed to variations in grain-boundary structure introduced during ceramic processing that can therefore exert an influence on total conductivity. These findings emphasize that compaction pressure can affect the electrochemical functionality through the impact on the sample microstructure and consequently on its impedance.

\newpage


\section{Conclusions}

In this work, the impedance spectroscopy of gadolinium doped ceria electrolytes were systematically investigated as a function of isostatic compaction pressure using in-house developed ceramic fabrication and processing tools. By isolating compaction pressure as the only variable during sample preparation, its influence on the impedance was identified.

The results demonstrate that changes in compaction pressure lead to measurable variations in impedance. This behavior cannot be explained only by intrinsic defect chemistry and is instead attributed to microstructural effects, particularly grain-boundary resistance.

Finally, a central hypothesis supported by the experimental data is that higher compaction pressures introduce microstructural changes that degrade the grain-boundary charge transport properties. These findings highlight the compaction pressure influence on electrochemical performance of GDC electrolytes. Future investigations  combining microstructural characterization with impedance analysis will be performed, and will further clarify the mechanisms linking the ceramic processing to the impedance behavior.

\vspace{60pt}


\funding{FAPESP process 24/01511-3 and 24/20599-9.}



\dataavailability{Data is provided within the manuscript.} 

\acknowledgments{The authors would like to thank the São Paulo Research Foundation (FAPESP) for its financial support (process 24/01511-3 and 24/20599-9).}


\isPreprints{}{
} 

\reftitle{References}


\bibliography{mybib}

\begin{thebibliography}{999}

\bibitem[Brett et~al.(2008)Brett, Atkinson, Brandon, and Skinnerd]{skinnerd:2008}
Brett, D.J.L.; Atkinson, A.; Brandon, N.P.; Skinnerd, S.J.
\newblock Intermediate temperature solid oxide fuel cells.
\newblock {\em Chem. Soc. Rev.} {\bf 2008}, {\em 37},~1568–1578.

\bibitem[Hussain and Yangping(2020)]{Saddam:2020}
Hussain, S.; Yangping, L.
\newblock Review of solid oxide fuel cell materials: cathode, anode,and electrolyte.
\newblock {\em Energy Transitions} {\bf 2020}, {\em 4},~113–126.

\bibitem[Li et~al.(2025)Li, Cai, and Amini~Horri]{GDC_LSGM_DATA}
Li, J.; Cai, Q.; Amini~Horri, B.
\newblock Highly conductive and stable electrolytes for solid oxide electrolysis and fuel cells: fabrication{,} characterisation{,} recent progress and challenges.
\newblock {\em Mater. Adv.} {\bf 2025}, {\em 6},~39--83.
\newblock {\url{https://doi.org/10.1039/D4MA00690A}}.

\bibitem[Acharya and Gaikwad(2022)]{Acharya17022022}
Acharya, S.A.; Gaikwad, V.M.
\newblock Understanding of dynamics of electrical processes in nanostructured Gd-doped ceria.
\newblock {\em Ferroelectrics} {\bf 2022}, {\em 588},~164--179,  \href{http://arxiv.org/abs/https://doi.org/10.1080/00150193.2022.2034451}{{\normalfont [https://doi.org/10.1080/00150193.2022.2034451]}}.
\newblock {\url{https://doi.org/10.1080/00150193.2022.2034451}}.

\bibitem[Li et~al.(2025)Li, Lu, Hou, Gao, and Li]{JiaHong2025}
Li, J.H.; Lu, F.F.; Hou, R.Y.; Gao, Y.; Li, C.X.
\newblock Elucidating the Sintering Mechanisms and Synergistic Doping Effects in CuO/Fe2O3 Codoped Gd-Doped Ceria Electrolytes for Advanced Low-Temperature Solid Oxide Fuel Cells (LT-SOFCs).
\newblock {\em ACS Applied Materials \& Interfaces} {\bf 2025}, {\em 17},~29813--29827,  \href{http://arxiv.org/abs/https://doi.org/10.1021/acsami.5c00238}{{\normalfont [https://doi.org/10.1021/acsami.5c00238]}}.
\newblock PMID: 40338612, {\url{https://doi.org/10.1021/acsami.5c00238}}.

\bibitem[Liang et~al.(2022)Liang, Tang, Zhou, Bai, Tian, Zhu, Zhou, Wang, and Yan]{Liang2022}
Liang, Q.; Tang, P.; Zhou, J.; Bai, J.; Tian, D.; Zhu, X.; Zhou, D.; Wang, N.; Yan, W.
\newblock Effect of MgO and Fe2O3 dual sintering aids on the microstructure and electrochemical performance of the solid state Gd0.2Ce0.8O2-$\delta$ electrolyte in intermediate-temperature solid oxide fuel cells.
\newblock {\em Frontiers in Chemistry} {\bf 2022}, {\em Volume 10 - 2022}.
\newblock {\url{https://doi.org/10.3389/fchem.2022.991922}}.

\bibitem[Liu et~al.(2025)Liu, Su, Li, Wang, Wang, Shang, and Xu]{Fangjie:2025}
Liu, F.; Su, Z.; Li, H.; Wang, Q.; Wang, X.; Shang, W.; Xu, B.
\newblock Status and prospects of electrode materials for symmetrical solid oxide fuel cells: a concise review.
\newblock {\em Ionics} {\bf 2025}, {\em 31},~3877--3894.

\bibitem[Antunes et~al.(2025)Antunes, de~Oliveira, de~Abreu, Dias, Brandão, Gonçalves, and and]{Unicamp2025}
Antunes, F.C.; de~Oliveira, J.P.; de~Abreu, R.S.; Dias, T.; Brandão, B.B.; Gonçalves, J.M.; and, J.R.
\newblock Reviewing metal supported solid oxide fuel cells for efficient electricity generation with biofuels for mobility.
\newblock {\em Journal of Energy Chemistry} {\bf 2025}, {\em 103},~106–153.

\bibitem[Gao et~al.(2016)Gao, Mogni, Miller, Railsback, and Barnett]{Liliana2016}
Gao, Z.; Mogni, L.V.; Miller, E.C.; Railsback, J.G.; Barnett, S.A.
\newblock A perspective on low-temperature solid oxide fuel cells.
\newblock {\em Energy Environ. Sci.} {\bf 2016}, {\em 9},~1602--1644.
\newblock {\url{https://doi.org/10.1039/C5EE03858H}}.

\bibitem[Vinchhi et~al.(2024)Vinchhi, Ray, Mallik, and Pati]{Vinchhi2024}
Vinchhi, P.; Ray, A.; Mallik, K.; Pati, R.
\newblock Gd-doped ceria with extraordinary oxygen-ion conductivity for low temperature solid oxide fuel cells.
\newblock {\em Scientific Reports} {\bf 2024}, {\em 14},~19010.
\newblock {\url{https://doi.org/https://doi.org/10.1038/s41598-024-59030-6}}.

\bibitem[Batista et~al.(2016)Batista, Ferreira, and Muccillo]{batista2016}
Batista, R.; Ferreira, A.; Muccillo, E.
\newblock Sintering and electrical conductivity of gadolinia-doped ceria.
\newblock {\em Ionics} {\bf 2016}, {\em 22},~1159–1166.
\newblock {\url{https://doi.org/https://doi.org/10.1007/s11581-016-1648-7}}.

\bibitem[Ghaemi et~al.(2024)Ghaemi, Sharifianjazi, Irandoost, Esmaeilkhanian, and {Amini Horri}]{GHAEMI2024}
Ghaemi, N.; Sharifianjazi, F.; Irandoost, M.; Esmaeilkhanian, A.; {Amini Horri}, B.
\newblock Electrical and microstructural characteristics of GDC electrolyte synthesised by benzoate coprecipitation for solid oxide fuel cells.
\newblock {\em Ceramics International} {\bf 2024}, {\em 50},~41780--41791.
\newblock {\url{https://doi.org/https://doi.org/10.1016/j.ceramint.2024.08.031}}.

\bibitem[Momin et~al.(2022)Momin, Manjanna, Aruna, {Senthil Kumar}, and Anjaneya]{MOMIN2022}
Momin, N.; Manjanna, J.; Aruna, S.; {Senthil Kumar}, S.; Anjaneya, K.
\newblock Effect of 20 mol 
\newblock {\em Ceramics International} {\bf 2022}, {\em 48},~35867--35873.
\newblock Advances in Ceramic Technologies: Materials and Manufacturing, {\url{https://doi.org/https://doi.org/10.1016/j.ceramint.2022.08.169}}.

\bibitem[Gupta et~al.(2022)Gupta, Shirbhate, Rambadey, Acharya, and Sagdeo]{MinalGupta2022}
Gupta, M.; Shirbhate, S.C.; Rambadey, O.V.; Acharya, S.A.; Sagdeo, P.R.
\newblock Temperature-Dependent Delocalization of Oxygen Vacancies in La-Substituted CeO2.
\newblock {\em ACS Applied Energy Materials} {\bf 2022}, {\em 5},~9759--9769,  \href{http://arxiv.org/abs/https://doi.org/10.1021/acsaem.2c01442}{{\normalfont [https://doi.org/10.1021/acsaem.2c01442]}}.
\newblock {\url{https://doi.org/10.1021/acsaem.2c01442}}.

\bibitem[Costilla-Aguilar et~al.(2021)Costilla-Aguilar, Pech-Canul, Escudero, Cienfuegos-Pelaes, and Aguilar-Martínez]{COSTILLAAGUILAR2021}
Costilla-Aguilar, S.; Pech-Canul, M.; Escudero, M.; Cienfuegos-Pelaes, R.; Aguilar-Martínez, J.
\newblock Gadolinium doped ceria nanostructured oxide for intermediate temperature solid oxide fuel cells.
\newblock {\em Journal of Alloys and Compounds} {\bf 2021}, {\em 878},~160444.
\newblock {\url{https://doi.org/https://doi.org/10.1016/j.jallcom.2021.160444}}.

\bibitem[Wang et~al.(2022)Wang, Jia, Lyu, Han, Wu, Sun, Iguchi, Yashiro, and Kawada]{WANG2022}
Wang, Y.; Jia, C.; Lyu, Z.; Han, M.; Wu, J.; Sun, Z.; Iguchi, F.; Yashiro, K.; Kawada, T.
\newblock Performance and stability analysis of SOFC containing thin and dense gadolinium-doped ceria interlayer sintered at low temperature.
\newblock {\em Journal of Materiomics} {\bf 2022}, {\em 8},~347--357.
\newblock {\url{https://doi.org/https://doi.org/10.1016/j.jmat.2021.09.001}}.

\bibitem[Cabrera-Pasca et~al.(2026)Cabrera-Pasca, Narváez-Romo, Ribeiro, Khan, {de Souza}, Britto-Costa, Ihringer, Figueroa, Martins, Khan, Machado, Meneghini, Brandon, and Lopes]{CABRERAPASCA2026}
Cabrera-Pasca, G.A.; Narváez-Romo, B.; Ribeiro, E.L.; Khan, L.U.; {de Souza}, M.B.; Britto-Costa, P.H.; Ihringer, R.; Figueroa, S.J.; Martins, T.M.; Khan, Z.U.;  et~al.
\newblock A study of a commercially available anode-supported 2R-Cell™ SOFC via synchrotron radiation XAFS and electrochemical characterization.
\newblock {\em Journal of Power Sources} {\bf 2026}, {\em 674},~239755.
\newblock {\url{https://doi.org/https://doi.org/10.1016/j.jpowsour.2026.239755}}.

\bibitem[Dias et~al.(2026)Dias, Antunes, {de Oliveira}, {dos Santos}, Soares, Doubek, and Zanin]{THIAGO2026}
Dias, T.; Antunes, F.C.; {de Oliveira}, J.P.; {dos Santos}, J.P.; Soares, D.; Doubek, G.; Zanin, H.
\newblock Influence of processing parameters on the electrical conductivity of 8YSZ electrolytes fabricated via tape casting.
\newblock {\em Progress in Engineering Science} {\bf 2026}, {\em 3},~100250.
\newblock {\url{https://doi.org/https://doi.org/10.1016/j.pes.2026.100250}}.

\bibitem[Zhou et~al.(2002)Zhou, Huebner, Kosacki, and Anderson]{Zhou2002}
Zhou, X.D.; Huebner, W.; Kosacki, I.; Anderson, H.U.
\newblock Microstructure and Grain-Boundary Effect on Electrical Properties of Gadolinium-Doped Ceria.
\newblock {\em Journal of the American Ceramic Society} {\bf 2002}, {\em 85},~1757--1762,  \href{http://arxiv.org/abs/https://ceramics.onlinelibrary.wiley.com/doi/pdf/10.1111/j.1151-2916.2002.tb00349.x}{{\normalfont [https://ceramics.onlinelibrary.wiley.com/doi/pdf/10.1111/j.1151-2916.2002.tb00349.x]}}.
\newblock {\url{https://doi.org/https://doi.org/10.1111/j.1151-2916.2002.tb00349.x}}.

\bibitem[LNN()]{LNNANOSITE}
\url{https://lnnano.cnpem.br/en/home-en/}.

\bibitem[Figueroa et~al.(2023)Figueroa, Rochet, Torquato, Espíndola, Jr., and Meyer]{Santiago2023}
Figueroa, S.J.; Rochet, A.; Torquato, I.F.; Espíndola, A.M.; Jr., H.R.; Meyer, B.C.
\newblock QUATI beamline: QUick x-ray Absorption spectroscopy for TIme and space-resolved experiments at the Brazilian Synchrotron Light Laboratory.
\newblock {\em Radiation Physics and Chemistry} {\bf 2023}, {\em 212},~111198.

\bibitem[Bowen et~al.(2020)Bowen, Johnson, McQuade, Wright, Kwong, Hsieh, Cann, and Woodside]{GDC_DATA1}
Bowen, M.S.; Johnson, M.; McQuade, R.; Wright, B.; Kwong, K.S.; Hsieh, P.Y.; Cann, D.P.; Woodside, C.R.
\newblock Electrical properties of gadolinia-doped ceria for electrodes for magnetohydrodynamic energy systems.
\newblock {\em SN Applied Sciences} {\bf 2020}, {\em 2},~1529.
\newblock {\url{https://doi.org/10.1007/s42452-020-03280-2}}.

\end{thebibliography}

%


\isPreprints{}{
} 
\end{document}